\documentclass[useAMS,usenatbib]{mn2e}
\usepackage{epsfig,amsmath,amssymb,amsfonts,mathrsfs,latexsym,graphicx}
\include{journals}
\newcommand{\eg}{e.g.\ }

\newcommand{\Msun}{{\rm M}_{\odot}}

\newcommand{\Nifs}{$^{56}$Ni}

\newcommand{\Mej}{$M_{\rm ej}$}
\newcommand{\KE}{$E_{\rm k}$}
\newcommand{\Eiso}{$E_{\rm iso}$}

\newcommand{\Eradio}{$E_{\rm radio}$}

\newcommand{\apj}{ApJ}
\newcommand{\apjl}{ApJL}

\newcommand{\aap}{A\&A}
\newcommand{\araa}{ARA\&A}
\newcommand{\mnras}{MNRAS}
\newcommand{\nat}{Nat}

\def\gsim{\mathrel{\rlap{\lower 4pt \hbox{\hskip 1pt $\sim$}}\raise 1pt \hbox {$>$}}}
\def\lsim{\mathrel{\rlap{\lower 4pt \hbox{\hskip 1pt $\sim$}}\raise 1pt \hbox {$<$}}}

\title[GRB/SNe are powered by magnetars]{An upper limit to the energy of 
gamma-ray bursts indicates that GRB/SNe are powered by magnetars}  

\author[P.A. Mazzali et al.]{
\parbox[t]{\textwidth}
{P. A. Mazzali$^{1,2,3}$\thanks{E-mail: P.Mazzali@ljmu.ac.uk},  
A.I. McFadyen$^4$, 
S.E. Woosley$^5$,
E. Pian$^6$, 
M. Tanaka$^7$
}
\vspace{6pt}
\\
$^1$Astrophysics Research Institute, Liverpool John Moores University,
146 Brownlow Hill, Liverpool L3 5RF, UK \\
$^2$INAF-Osservatorio Astronomico, vicolo dell'Osservatorio, 5, I-35122 Padova, Italy \\
$^3$Max-Planck Institut f\"ur Astrophysik, Karl-Schwarzschildstr. 1, D-85748 
Garching, Germany \\
$^4$Physics Dept., New York University, New York, NY, USA \\
$^5$Astronomy Dept., University of California, Santa Cruz, Santa Cruz, Ca, USA\\
$^6$INAF IASF,  Via P. Gobetti 101, I-40129 Bologna, Italy\\
$^7$National Astronomical Observatory of Japan, Mitaka, Tokyo, Japan \\
}

\begin{document}

\date{Accepted ... Received ...; in original form ...}

\pagerange{\pageref{firstpage}--\pageref{lastpage}} \pubyear{2014}

\maketitle

\label{firstpage}


\begin{abstract}
The kinetic energy of supernovae (SNe) accompanied by gamma-ray bursts (GRBs)
tends to cluster near $10^{52}$\,erg, with $2 \times 10^{52}$\,erg an upper 
limit to which no compelling exceptions are found (assuming a certain degree of
asphericity), and it is always significantly larger than the intrinsic energy of
the GRB themselves (corrected for jet collimation). This energy is strikingly
similar to the maximum rotational energy of a neutron star rotating with period
1 ms. It is therefore proposed that all GRBs associated with luminous SNe are 
produced by magnetars. GRBs that result from black hole formation (collapsars) 
may not produce luminous SNe.
X-ray Flashes (XRFs), which are associated with less energetic SNe, 
are produced by neutron stars with weaker magnetic field or lower spin. 
\end{abstract}



\section{Introduction}

The connection between long-duration Gamma-Ray Bursts (GRBs) and SNe is well
established in the local Universe. Following the first cases of coincidences
\citep{galama1998,Stanek2003,Hjorth2003,matheson2003,malesani2004,galyam2004,thomsen2004},
now for almost every GRB at redshift $z \lsim 0.3$ a corresponding SN has been
identified 
\citep{woosleybloom2006,pian2006,mirabal2006,modjaz2006,chornock2010,cano2011a,bufano2012,olivares2012,melandri2012,xu2013,melandri2014}. 
Despite the diverse properties of their associated GRBs, all GRB/SNe observed so
far are luminous, broad-lined Type Ic SNe \citep[no H, no He,][]{filipp97}.  
The very broad lines indicate a high expansion velocity of the ejecta, and point
to a high explosion kinetic energy (\KE) \citep{mazzali2000}.  

Detailed models of GRB/SNe yield a typical SN \KE\ of a few $10^{52}$\,erg
(depending on the asphericity of the SN), an ejected mass \Mej\,$\sim 10 \Msun$,
and a \Nifs\ mass of $\sim 0.4 \Msun$.  This places GRB/SNe at the luminous,
energetic and massive end of SNe\,Ic \citep[\eg][]{mazzali2013} and points to a
massive star origin \citep[\eg][]{mazzali2006a}.  
Two recent events confirm and reinforce this trend: SN\,2013cq/GRB130427A
\citep{maselli2014} and SN2013dx/GRB130702A \citep{singer2013}. Although the two
GRBs are very different in energy, the former extremely powerful, similar to
cosmological ones, the latter a normal GRB, the SNe are again similar
\citep[][D'Elia et al., in prep.]{melandri2014}.

It has been proposed that long GRBs are produced by the Collapsar mechanism,
where a massive star collapses directly to a black hole (BH). Accretion on the BH
releases energy in the form of a relativistic jet which may explode the star and
produce a visible SN if \Nifs\ is synthesised in the dense neutrino wind
emanating from the accretion disc \citep{woosley1993,mcfw1999}. 

SNe\,Ic associated with X-ray Flashes (XRFs) have a smaller \KE, more similar to
ordinary SNe\,Ic, and are not as luminous
\citep{pian2006,mazzali2006b,bufano2012}.  Models indicate progenitor stars of
$\sim 20-25 \Msun$, which are expected to collapse to Neutron Stars (NS). Their
\KE\ (a few $10^{51}$\, erg) appears to be consistent with energy injection from
a magnetar, a rapidly spinning magnetised NS \citep{mazzali2006b}. This
mechanism taps the energy in the magnetic field and may also give rise to a
relativistic jet \citep[see e.g.,][]{thompson2004,dessart2008}.  

Observational and theoretical evidence has been mounting that more massive stars
can also collapse to NS \citep{Muno06,ugliano2012,sukhwoos14}.  Magnetar jets
and their potential as a source of GRBs have been investigated in various
papers, suggesting that magnetar energy can be used to energise GRBs or XRFs
\citep{bucciantini2007,bucciantini2008,bucciantini2009,Metzger2011}. 

It has also been proposed that very rapidly spinning magnetars can explain the
much brighter light curves of GRB/SNe \citep{Umcf2007}. This may conflict with
the observation in SN\,1998bw of strong emission lines of Fe, which indicate a
high \Nifs\ yield \citep{patat2001,mazzali2001}.  On the other hand, only
SN\,1998bw could be followed late enough to observe Fe lines. 

One of the most interesting unsolved questions in GRB science is what actually
drives the event. In the collapsar model the jet generated by the BH explodes
the star, but is its energy sufficient to impart a high \KE\  to the SN?
Simulations have so far not tested this, but the energy needed for the jet to
emerge from the star and unbind it  
\citep[$\sim 3 \times 10^{51}$\,erg,][]{lazzati2013} appears to be much smaller
than the SN \KE. In the magnetar scenario, if the coupling is large energy may
be extracted from the NS and added to the SN \KE, which would otherwise derive
from the classical neutrino  mechanism. The sub-relativistic outflow may not be
highly collimated, as indicated by the distribution of SN material
\citep{mazzali2006b,mazzali2007}. In this scenario energy production would be
limited by the NS spin rate. 

We analyse the global properties of the GRBs and their SNe in order to look for 
indications of a preferred mechanism. We compare the energies of GRBs, XRFs, and
their accompanying SNe. In Section 2.1 we estimate the intrinsic energy of
low-redshift GRBs ($z \lsim 0.35$) with associated SNe by applying a correction
for the jet opening angle to the observed $\gamma$-ray energies. In Section 2.2
we estimate the energy in relativistic ejecta as probed by radio data. In
Section 2.3 we compare both of these to the SN \KE\ as derived from modelling. 
In Section 3 we present our results.  In Section 4 we extend the comparison to
all GRBs at higher redshift for which a SN was reported and discuss our
findings.

\section{Nearby GRBs, XRFs, and their SNe}

Isotropic-equivalent energies (\Eiso) of nearby GRBs connected with
well-studied SNe are extremely diverse. GRB\,980425 had a very low \Eiso,
which was one of the aspects that raised doubts on the reality of the first
GRB/SN association.  On the other hand GRB030329, associated with SN\,2003dh,
was similar to many long GRBs. GRB130427A has \Eiso $\sim 10^{54}$\,erg,
comparable to cosmological GRBs.  However, \Eiso is unlikely to be the real
jet energy.

The true energy of the jet, $E_{\gamma}$, can be estimated from \Eiso,
adopting a correction for collimation. Alternatively, radio energy is thought to
be a good proxy for the energy of relativistic material, assuming that this
energy is completely used up in the interaction with circumstellar
material and radiated isotropically at later times (jet radio calorimetry). 

A model-dependent estimate of $E_{\gamma}$ can be obtained from the timing of
the break in the afterglow light curve. An achromatic break may indicate that
the edge of the jet swept past our viewing point. This information is however
not always available. Its absence may indicate lack of collimation but also just
be due to incomplete data. Once \Eiso has been corrected for jet
collimation, which can be quite uncertain \citep[see e.g. ][]{cenko2010}, it can
be compared with the SN \KE\ and with the radio energies.


{\small
\begin{table*}
\begin{center}
  \caption{Properties of GRB/SNe at $z \lsim 0.3$.}
  \begin{tabular}{cccccccccc}
  \hline
  GRB/SN   	 &  $z$   &  T90   &   \Eiso  & $\theta_{op}$ &  $E_{\gamma}$  &   SN \KE\      &    M($^{56}$Ni)&    \Eradio   &  Refs. \\ 
                 &        &   [s]  &  [$10^{50}$ erg] &   [deg]   &   [$10^{50}$ erg] & [$10^{50}$ erg]&   [$M_\odot$]  &  [$10^{50}$ erg] &  \\
        1        &    2   &    3   &        4         &     5     &         6         &        7       &       8        &         9        & 10 \\
 \hline
980425  / 1998bw & 0.0085 &   30 & $0.010 \pm 0.002$ & 180        & $0.010 \pm 0.002$ & $500 \pm  50$ & $0.43 \pm 0.05$ & $\sim$0.2        & 1-3 \\
030329  / 2003dh & 0.1685 &   23 & $  150 \pm 30$    & $6 \pm 2$  & $0.23 \pm  0.05$  & $400 \pm 100$ & $0.4 \pm 0.1$   & $2.5 \pm 0.8$    & 2,4-7 \\
031203  / 2003lw & 0.1055 &   40 & $  1.0 \pm 0.4$   & 180        & $1.0 \pm 0.4$     & $600 \pm 100$ & $0.6 \pm 0.1$   & $0.17 \pm 0.06$  & 1,2,8 \\
060218  / 2006aj & 0.0335 & 2000 & $ 0.53 \pm 0.03$  & 180        & $0.53 \pm 0.03$   & $20 \pm 6$ & $0.20 \pm 0.05$ & $0.020 \pm 0.006$ & 4,9-11 \\
100316D / 2010bh & 0.059  & $>$1300 & $ 0.7 \pm 0.2$ & 180        & $0.7 \pm  0.2$    & $100 \pm  60$ & $0.12 \pm 0.02$ & $\sim$0.2        & 12-14 \\ 
120422A / 2012bz & 0.283  &    5 & $ 2.4  \pm 0.8$   & $23 \pm 7$ & $0.05  \pm 0.02$  & $400 \pm 100$ & $0.3 \pm  0.1$  &  --              & 15,16 \\
130427A / 2013cq & 0.3399 &  160 & $8100  \pm 800$   & $3 \pm 1$  & $4 \pm 1$     & $640 \pm  70$   & $0.4 \pm 0.1$ & $6 \pm 2$          & 17-20 \\
130702A / 2013dx & 0.145  &   59 & $ 6.5 \pm  1.0^a$ & $14 \pm 4$ & $0.05 \pm 0.02$   & $300 \pm  60$ & $0.3 \pm 0.1$   & $20 \pm 5$       & 21-23 \\
\hline
\hline
\multicolumn{10}{l}{$^a$  The uncertainty on \Eiso is here amended (L. Amati, priv. comm.)}\\
\multicolumn{10}{l}{References: 1. Amati (2006); 2. Mazzali et al. (2006a);  
 3. Li \& Chevalier (1999); 4. Amati et al. (2008); 5. Deng et al. (2005);}\\
\multicolumn{10}{l}{6. Berger et al. (2003); 7. Gorosabel et al. (2006); 
 8. Soderberg et al. (2004); 9. Mazzali et al. (2006b); 10. Pian et al. (2006);} \\
\multicolumn{10}{l}{11. Soderberg et al. (2006); 12. Starling et al. (2011); 
 13. Bufano et al. (2012); 14. Margutti et al. (2013); 15. Melandri et al. (2012);}\\
\multicolumn{10}{l}{16. Schulze et al. (2014); 17. Maselli et al. (2014); 
 18. Melandri et al. (2014); 19. Xu et al. (2013); 20. Perley et al. (2013);}\\
\multicolumn{10}{l}{21. Singer et al. (2013); 22. D'Elia et al., in prep.; 
 23. Amati et al. (2013).}\\

\end{tabular}
\end{center}
\end{table*}

}


\subsection{Gamma-ray energies of GRBs}

Values of \Eiso\ of GRB/SNe 
are listed in Table 1. If an estimate of the jet opening angle $\theta_{op}$ was
available in the literature or could be derived from the afterglow
multiwavelength light curves \citep[as outlined by][]{sari1999}, we reported
this angle and computed the collimation-corrected energy $E_{\gamma}$. 


The optical light curves of GRB130702A steepen at $t_{obs}=1.17$ days, and the
X-ray light curve is compatible with a steepening at about the same time
\citep{singer2013}.  If that is a jet break, the jet opening angle is $\sim$14
degrees and $E_{\gamma} \sim 5 \times 10^{48}$ erg.  A SN similar to SN\,1998bw
was indeed detected in coincidence with GRB130702A (D'Elia et al., in prep.). No
correction is possible for GRBs 980425 and 031203 or the two XRFs.

\subsection{Radio energies}

We list GRB/SN energies from radio measurements (\Eradio) in Table 1. 
Whenever possible, we took estimates from the literature. For
GRB130427A/SN2013cq and GRB130702A/SN2013dx \Eradio was estimated from the
available radio data following \citet{lichevalier1999}.  In the case of
GRB130702A/SN2013dx there may be a significant contribution from the afterglow,
because radio measurements were taken only 2 rest-frame days after the
explosion.  Since it is impossible to disentangle the contribution to the radio
emission by the GRB from that of the SN, these values must be regarded as upper
limits to the SN energy.  No radio observations are available for
GRB120422A/SN2012bz.

\subsection{Kinetic energies of nearby GRB/SNe}

We obtained SN \KE\ (Table 1) from models or from spectroscopic analogues.  In
particular, \KE\ of SNe\,1998bw, 2003dh and 2003lw are from the re-analysis of
\citet{mazzali2006a}.  For SN\,2012bz \KE\ is taken to be equal to that of
SN\,2003dh based on the similarity of the spectra and the light curve
\citep{melandri2012}; M(\Nifs) is only 15\% less than for SN\,2003dh.  For
SN\,2013cq we used the \KE\ estimate of \citet{xu2013}, who find a value similar
to that of SN1998bw. This is also supported by the fact that the bolometric
light curve maximum of SN\,2013cq, which was accurately measured with HST
\citep{levan2013,melandri2014}, is consistent with that of SN1998bw.  For
SN\,2013dx, whose light curve is similar to those of other GRB/SNe (D'Elia et
al. 2014, in preparation), we took the average of the \KE\ of SN\,2010ah
\citep{mazzali2013} and SN\,1998bw based on the spectroscopic similarity.  For
SN\,2006aj \KE\ was obtained through modelling \citep{mazzali2006b}. For
SN\,2010bh it was estimated by \cite{bufano2012}. 

All these \KE\ assume spherical symmetry. However, we know from the distribution
of elements (in particular Fe and O as observed through their nebular emission
lines) that at least SN\,1998bw was significantly aspherical, and was observed 
near the direction of most rapid expansion, which is consistent with the
detection of the GRB \citep{mazzali2001}. Therefore, spherically symmetric \KE\
are likely to be overestimated.  Using 2D explosion models and 3D radiation
transport calculations, \citet{maeda2002} and \citet{tanaka2007} found that the
real SN \KE\ may be a factor of 2-5 smaller.  This correction, which is not
shown in Fig. 1, would cause the six GRB/SNe to cluster around \KE\,$\sim (1-2)
\times 10^{52}$\,erg.  The \KE\ of the two XRF/SNe do not require a correction,
because there is no evidence for asymmetry in SN\,2006aj \citep{mazzali2007}.

\section{Results}

We can now compare the various energies. 

In Fig. 1a the collimation-corrected GRB energy, $E_{\gamma}$, is compared to
the SN \KE, not corrected for asphericity. GRB \Eiso values range over 6
orders of magnitude. $E_{\gamma}$ values still cover 3 orders of magnitude, and
are always significantly smaller than any SN \KE\  (the diagonal line is \KE\ $=
E_{\gamma}$). This suggests that the GRB jet is unlikely to be the driving
phenomenon behind GRB/SNe, as the SN carries most of the energy, as already
noted by \citet{woosleybloom2006}. 

Fig. 1b shows \Eradio vs the SN \KE.  Again, the SN energy is always much
larger.

Finally, Fig. 1c shows the \Eradio v. $E_{\gamma}$. The GRB energies
estimated from the jet break and from the radio, which rely on different
wavebands and on observations taken at completely different times, are in
general agreement, but the SN \KE\ are much larger than both. This confirms that
\Eradio is a good proxy for $E_{\gamma}$, but not for either the SN \KE\
or the total energy of the event.


\begin{figure*}

\begin{tabular}{ccc}

\begin{minipage}{0.32\textwidth}
  \includegraphics[width=1.38\textwidth]{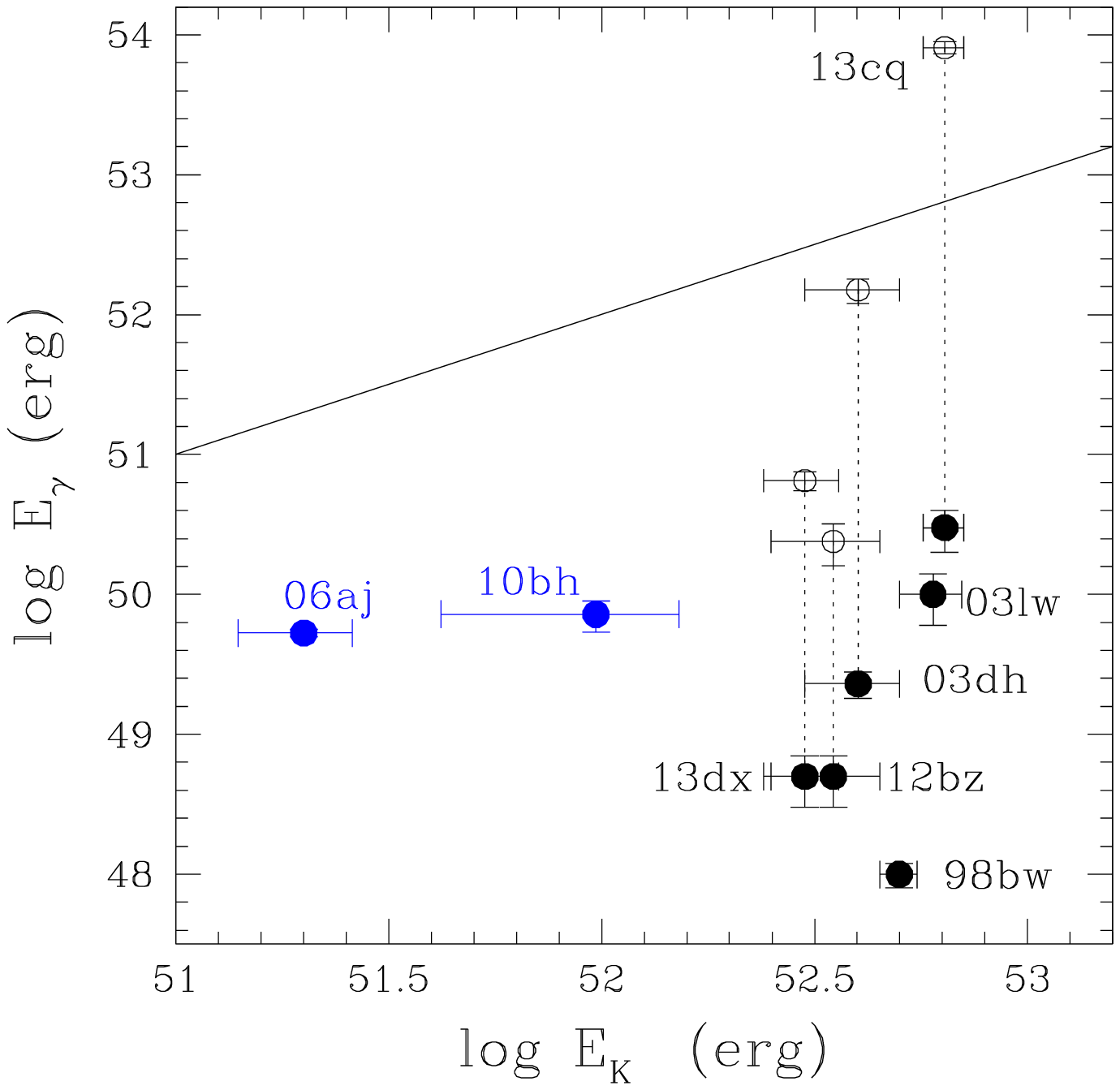}
\end{minipage} 
&
\begin{minipage}{0.32\textwidth}
  \includegraphics[width=1.38\textwidth]{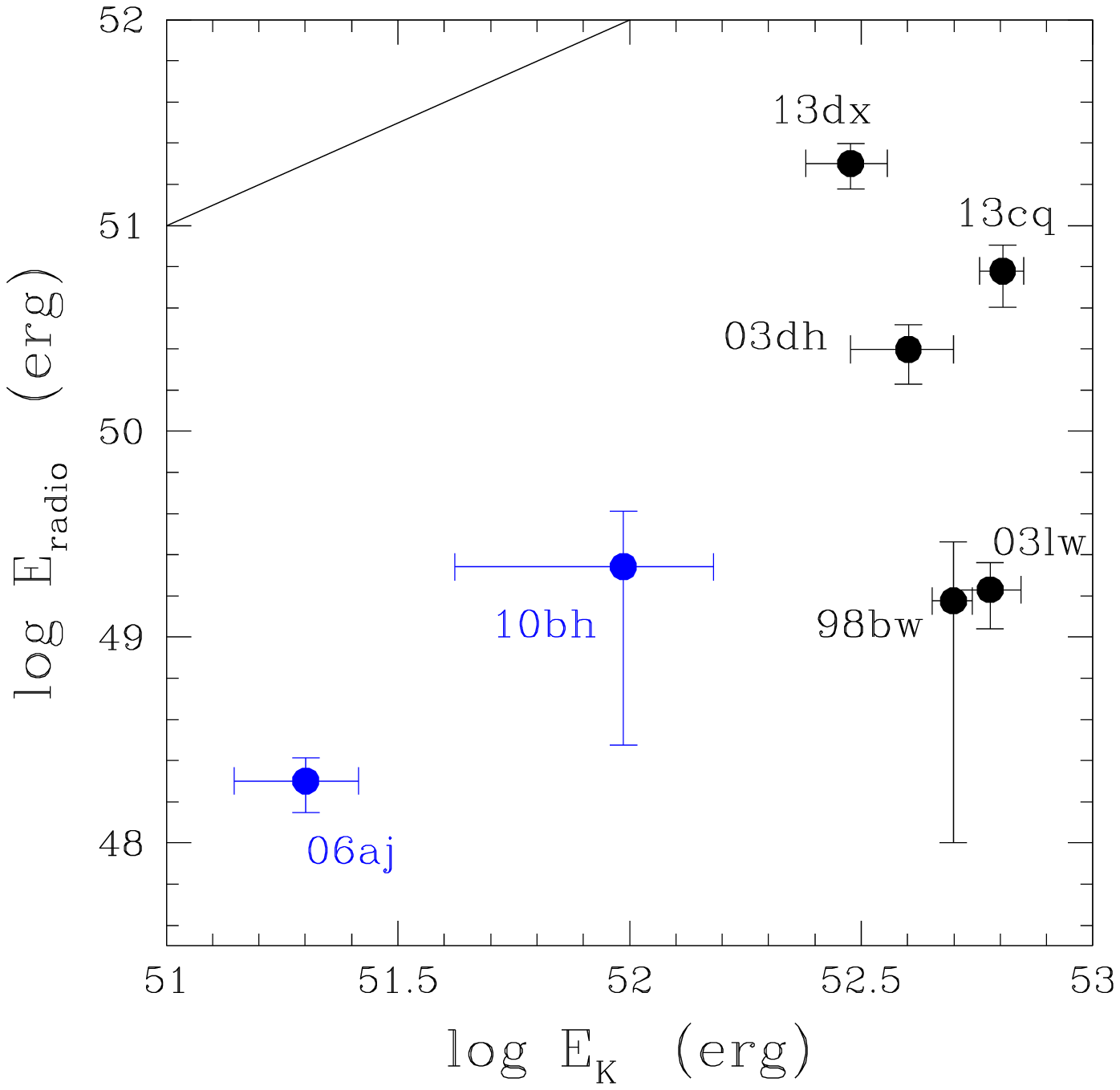}
\end{minipage}
&
\begin{minipage}{0.32\textwidth}
  \includegraphics[width=1.38\textwidth]{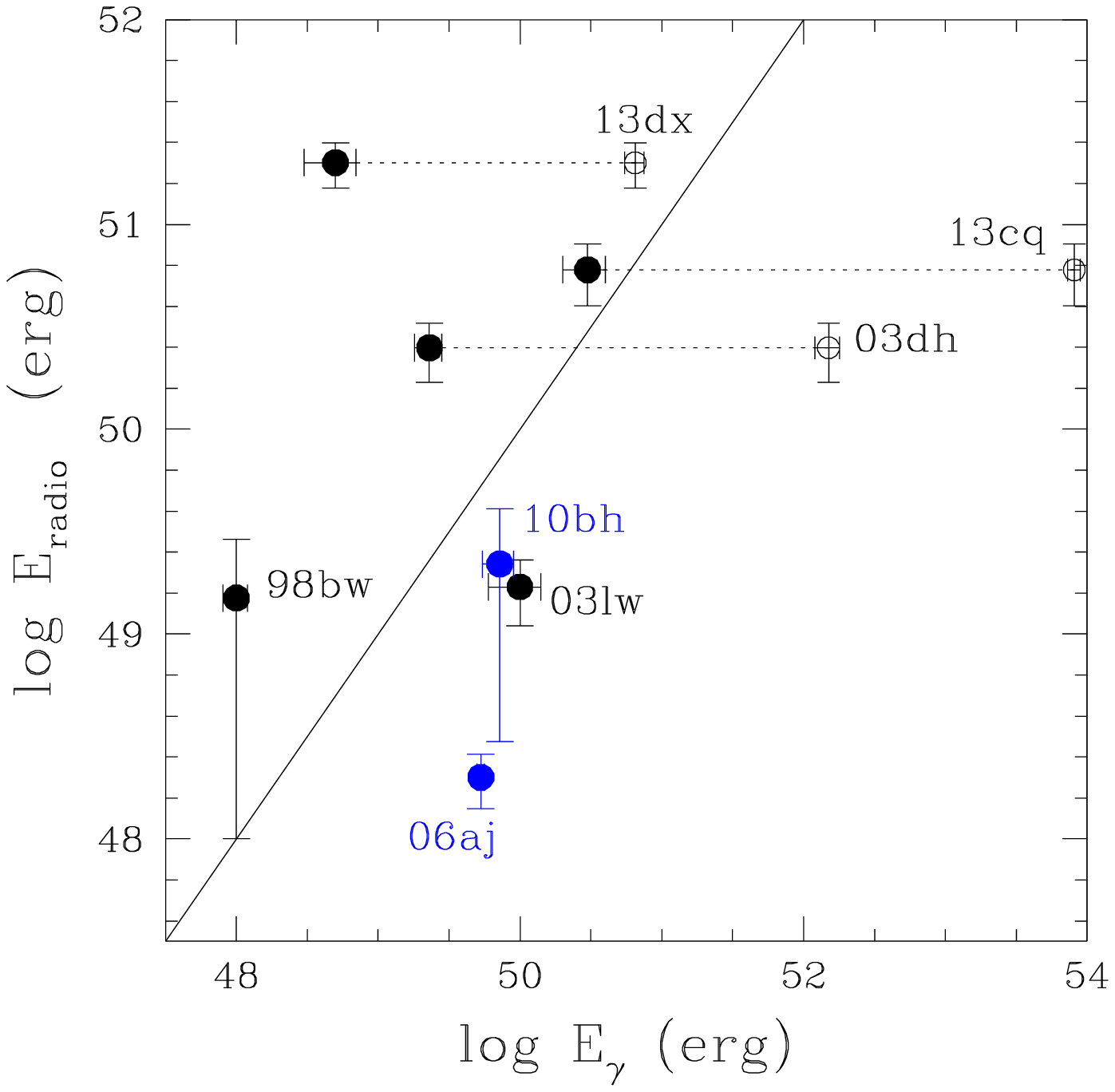}
\end{minipage}

\end{tabular}

\caption{Mutual dependences of the $\gamma$-ray energies emitted in the prompt
GRB and XRF events, the isotropic kinetic energies of the associated SNe, \KE,
and the energies inferred from radio observations \Eradio.  Note that in (a)
we have arbitrarily lowered the \KE of SN\,2012bz from $4 \times 10^{52}$ to
$3.5 \times 10^{52}$  erg, to avoid overlap with SN\,2003dh.  In (a) and (c)
when an estimate of the opening angle of the GRB jet exists,
isotropic-equivalent energies \Eiso are shown as open circles, and connected
with dashed lines to the corresponding values of the collimation-corrected
energies $E_{\gamma}$, shown as filled circles.   Black symbols are GRBs and
blue symbols are XRFs. } 

\end{figure*}


\section{Discussion}
 
Well-studied GRB/SNe have a roughly standard energy, \KE\,$\sim$\,(1-2)$\times
10^{52}$\,erg, if account is taken for asphericity. XRF/SNe have a smaller \KE\
by about a factor of 10. All SN \KE\ are much larger than all of the GRB/XRF 
$E_{\gamma}$, which seem to be capped at a few $10^{50}$\,erg.  This suggests
that the GRB/SN phenomenon is driven by the SN, not the GRB jet. This evidence
challenges a picture in which the relativistic jet explodes the star.
Simulations \citep{lazzati2013} have shown that the jet can unbind the star, but
it is not clear how the jet can transfer the required large \KE\ to the star
once it has escaped without $E_{\gamma}$ increasing to unreasonably large
values. Also, it may be less natural for GRB jets to produce SNe with always the
same total \KE\ and the same amount of \Nifs\ to such a degree
\citep{melandri2014}. 

On the other hand, the energetics of GRB/SNe are strikingly similar to the
rotational energy of a millisecond magnetar. The rotational energy of a
rapidly rotating NS is $E_{rot} \sim 2 \times 10^{52} (M/1.4 \Msun) (R/10{\rm
km})^2 / (P/1{\rm ms})^2$, where $M$ is the NS mass, $R$ its radius and $P$ its
spin period. Rotational energy can be tapped by rapid spindown on a GRB
timescale if the magnetic field is in the magnetar range \citep[$B \sim 10^{15}$
G;][]{Usov92,duncanthom1992}. 

This leads us to propose that in GRB/SNe the exploding star gives birth to a
highly magnetized millisecond NS. Deposited magnetar energy can further energize
the SN, and \KE\ $\sim 10^{52}$\,erg is a limit to the total intrinsic energy of
GRB/SNe. This limit has not been violated by any GRB/SN so far, if they are all
aspherical.  

Magnetar outflows can be focussed into magnetic jets by interaction with the
stellar envelope because hoop stress tends to collimate the flow after it comes
into pressure equilibrium with the shocked stellar cavity from which the
magnetar formed. The collimated magnetic wind (which is sometimes called a
"magnetic tower" and is similar to the collimation of pulsar wind nebulae such
as the Crab nebula) can burrow its way out of the star.  A very small fraction
of the total energy is seen to emerge in the relativistic jet. If a large
fraction of the magnetar energy can be transferred to the progenitor star,
mostly near the jet axis \citep{bucciantini2009}, it can be added to the SN
energy. The energy deposited also contributes to increasing the isotropic
component of the SN \KE\ \citep{mazzali2006b}. The SN can take on an
increasingly aspherical shape the higher the energy contribution from the
magnetar \citep[GRB/SNe are more aspherical than XRF/SNe, ][]{mazzali2007}. 

In this scenario, \Nifs\ may be produced as the expanding magnetar wind shocks
the inner star. If this happens quasi-spherically, before the star expands too
much, sufficient material can be shocked to produce the several $0.1 \Msun$ of
\Nifs\ in an almost spherical distribution required by GRB/SN light curves
\citep{maeda2003}. The collimated magnetic wind may produce some more \Nifs\ at
high velocities, as also required by the rapid rise of GRB/SN light curves.  The
late-time deposition of magnetar energy may also contribute to the SN light
curve, along with \Nifs. Late-time spectra of more GRB/SNe would be necessary to
clarify how much \Nifs\ is actually produced through the observation of emission
lines of Fe. Presently this information is only available for the nearest event,
SN\,1998bw \citep{mazzali2001}. 

The range of GRB prompt emission energy could be produced by interaction of the
jet as it propagates through the stellar envelope. A range of several orders of
magnitude in $E_{\gamma}$ may be possible, since the jet may be slowed down to
variable degrees by the development of instabilities or by interaction with
extended outer layers of the star. Small amounts of baryons mixed into the jet
can ``pollute" it and reduce its $\gamma$-ray luminosity. An extended envelope
may even block the jet altogether \citep{mazzali2008}.

Magnetars have been proposed to energise X-ray Flashes (XRF) and their
associated SNe\,Ic \citep{mazzali2006b}.  XRF/SNe have less extreme properties
than GRB/SNe, in particular they have smaller \KE\  (a few $10^{51}$\,erg),
luminosities [M(\Nifs)\,$\sim 0.2 \Msun$, only marginally larger than in
ordinary core-collapse SNe], and progenitor masses  
\citep[$\sim 20\Msun$,][]{mazzali2006b}.   They are less aspherical than 
GRB/SNe \citep{mazzali2007}.  They may be the result of lower-spin magnetars. 

The progenitors of GRB/SNe are thought to be stars of  M$_{\rm ZAMS} \sim 30-50
\Msun$. If GRB/SNe are also powered by magnetars then at least some of these
stars also collapse to NS.\footnote{Since estimates of the mass of GRB/SN
progenitors \citep[\eg][]{mazzali2013} are based on removing a BH remnant of
typically $3\Msun$, if the remnant is a NS instead masses may have to be revised
downwards slightly.}  Since GRBs and XRFs exhibit a continuum of properties,
this picture reconciles their appearance with their origin as a single
mechanism.  Indeed, \citet{burrdess07} find that jets are always produced 
when a proto-NS is formed, if the magnetic field is very high.

Direct collapse to a BH may not necessarily lead to a luminous SN. The \Nifs\
produced by the disk wind \citep{mcfw1999} could be highly variable and may
accrete into the BH, in the spirit of the initial proposal of a ``failed SN"
\citep{woosley1993}. This may be the case of the 2 low-redshift GRBs, 060614 and
060505, which showed no SN down to M(\Nifs) $ \sim 0.01 M_\odot$
\citep{mdv2006,fynbo2006,galyam2006,ofek2007}.  Fallback of \Nifs\ onto the BH
is one possibility \citep{Moriya2010}. On the other hand, both of these GRBs
have $E_{\gamma}$ well below the magnetar limit. 


Magnetars have also been proposed as the energy source for GRBs
\citep{thompson2004}, for GRB/SNe and luminous SNe\,Ib/c
\citep{KasenBildsten2010,woosley2010} and for the peculiar SN\,Ib 2005bf
\citep{maeda2007}. 
\cite{luzhang2014} find that most GRBs are compatible with being energised by a
magnetar. On the other hand, \cite{cenko2010} find that the energetics of three
out of five well-observed high-redshift Swift GRBs have energies similar to the
maximum energy provided by a spinning NS ($10^{52}$\,erg) even after correction
for collimation.  Only one of these, GRB080319B, may show a bump in its light
curve, but any SN would be very faint \citep{tanvir2010}. These may indeed all
be collapsars. 

We checked all other GRBs, at any redshift, for which a SN was reported.  Their
\Eiso\ almost ever exceeds a few $10^{52}$ erg. Five (991208, 000911, 011121,
020405, 090618) have \Eiso\,$\sim 10^{53}-10^{54}$\,erg
\citep{amati2008,2009gcn9530}, but in four of these the optical afterglows
exhibit potential jet breaks, leading to substantial energy collimation
corrections \citep{ajct2001,greiner2003,price2003,cano2011a}. No breaks are
reported in the optical afterglow of GRB000911 \citep{lazzati2001,masetti2005},
but the light curve could admit a break at t$\sim$5 rest-frame days, leading 
to an energy collimation correction factor of at least 100. 
For GRB111211 \citep{lazzarotto2011,deugarte2012}, no $\gamma$-ray energy or 
fluence is available.  From the GRB peak flux ($1.5 \times 10^{-6}$
erg\,s$^{-1}$\,cm$^{-2}$ in 20-60 keV), duration (15 s) and redshift  
\citep[$z = 0.478$,][]{vergani2011} we estimate \Eiso $\lsim 10^{52}$\,erg.  
All these events may be driven by magnetars.


\section*{Acknowledgments} 

We thank the Yukawa Institute of Kyoto University and the National Astronomical
Observatory of Japan for the hospitality while some of this work was carried
out.  EP acknowledges support from ASI INAF I/088/06/0, INAF PRIN 2011 and PRIN
MIUR 2010/2011. SEW has been supported by the NASA Theory Program (NNX09AK36G)
and the High Energy Physics Program of the DOE (DE-SC0010676).



\bsp

\label{lastpage}

\end{document}